\documentclass[twocolumn,aps,prl,superscriptaddress,floatfix,longbibliography,10pt]{revtex4-1}

\usepackage[latin9]{inputenc}
\usepackage{amsmath}
\usepackage{amssymb}
\usepackage{graphicx}
\usepackage{upgreek}
\usepackage{dsfont}
\usepackage{xcolor}
\usepackage{booktabs} 
\usepackage{multirow}
\usepackage[colorlinks=true,citecolor=blue,linkcolor=blue]{hyperref}

\newcommand{\vect}[1]{\boldsymbol{#1}} 

\newcommand{\mathcomma}{~ ,}       
\newcommand{\mathperiod}{~ .}           
\newcommand{\dosf}{\rho_{\mathrm{F}}}     

\DeclareMathOperator{\tr}{tr}
\newcommand{\dd}{\mathrm{d}}

\begin{document} 
\title{Role of fluctuations for density-wave instabilities:  Failure of the mean-field description}
\author{Mareike Hoyer}
\affiliation{Institut für Theorie der Kondensierten Materie, Karlsruher Institut für Technologie, D-76131 Karlsruhe, Germany}
\affiliation{Institut für Festkörperphysik, Karlsruher Institut für Technologie, D-76021 Karlsruhe, Germany}
\author{Jörg Schmalian}
\affiliation{Institut für Theorie der Kondensierten Materie, Karlsruher Institut für Technologie, D-76131 Karlsruhe, Germany}
\affiliation{Institut für Festkörperphysik, Karlsruher Institut für Technologie, D-76021 Karlsruhe, Germany}
\date{\today}
\begin{abstract}
Density-wave instabilities have been observed and studied in a multitude of materials. 
Most recently, in the context of unconventional superconductors like the iron-based superconductors, 
they have excited considerable interest. 
We analyze the fluctuation corrections to the equation of state of the density-wave order parameter for commensurate charge-density waves and spin-density waves due to perfect nesting. For XY~magnets, we find that  contributions due to longitudinal and transverse fluctuations cancel each other, making the mean-field analysis of the problem controlled. This is consistent with the analysis of fluctuation corrections to the BCS~theory of superconductivity [\v{S}.\ Kos, A.\ J.\ Millis, and A.\ I.\ Larkin, Phys.~Rev.~B~\textbf{70}, 214531 (2004)]. However, this cancellation does not occur in density-wave systems when the order parameter is a real $N$-component object with $N\neq2$. Then, the number of transverse fluctuating modes differs from the number of longitudinal fluctuating modes. Indeed, in the case of charge-density waves as well as spin-density waves with Heisenberg symmetry, we find that fluctuation corrections are not negligible, and hence mean-field theories are not justified.  These singular fluctuations originate from the intermediate length-scale regime, with wavelengths between the lattice constant and the $T=0$ correlation length. 
The resulting logarithmic fluctuation contributions to the gap equation originate from the derivative of the anomalous polarization function, and the crucial process is an interaction of quasiparticles through the exchange of fluctuations.
\end{abstract}
\maketitle
\section{Introduction}
\begin{figure}[t]
\includegraphics[width=\columnwidth]{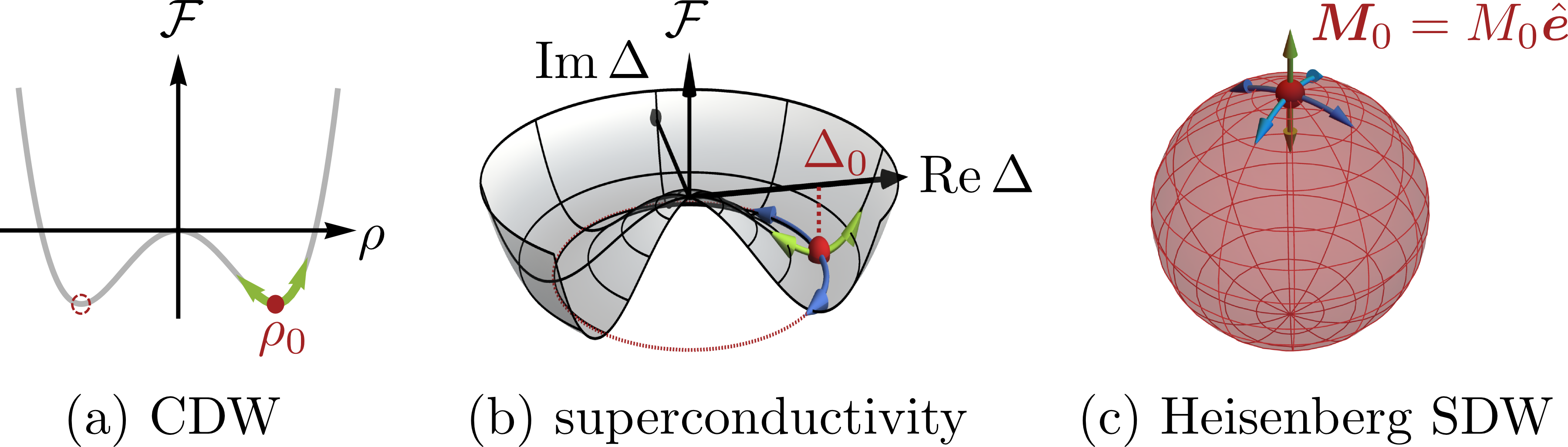}
\caption[Visualization of fluctuations of the order parameter]{Visualization of the order parameter and fluctuations around its mean-field value (red dot) for three examples: (a) commensurate charge-density wave order, (b) superconductivity, and (c) commensurate spin-density wave order of Heisenberg spins. The longitudinal mode is indicated in green while the $N-1$~transverse modes are shown in blue. In case of (a) and (b), the manifold of degenerate ground states (red) is shown within the corresponding energy landscape (gray) below the phase transition, whereas for (c) only the former.}
\label{fig:visu-fluctuations}
\end{figure}

In condensed matter systems, the interaction between electrons often gives rise to an instability of the Fermi-liquid ground state, resulting in the formation of a low-temperature ordered phase. Examples of such ground states include superconductivity, charge-density wave order and magnetically ordered states, and are expected in all materials at least in the clean limit. 
Often, the low-temperature ordered phase can be described in terms of an effective mean field that characterizes the novel ground state and serves as an order parameter (Fig.~\ref{fig:visu-fluctuations}). 
The poster child of such a mean-field theory is the celebrated BCS~theory~\cite{BardeenCooperSchrieffer-PR106.1957,BardeenCooperSchrieffer-PR108.1957} of superconductivity, which offers arguably the single most successful mean-field description of an interacting many-body system. The principal reasons for the success of this theory are the following: 
\begin{enumerate}
 \item[(i)] the instability towards the BCS~state occurs for arbitrarily small interactions,  
 \item[(ii)] fluctuations beyond mean-field theory can be neglected for small interaction strength since the coherence length is large compared to the inverse Fermi momentum in conventional superconductors, $\xi\gg  k_\mathrm{F}^{-1}$, resulting in a narrow Ginzburg regime. 
\end{enumerate}
There are instabilities in the particle-hole channel with perfect nesting~\cite{Peierls1955,Overhauser-PR1962} which share the first characteristic with the BCS~theory~\cite{FeddersMartin-PR1966,ChanHeine-JPhysF1973,Gorkov-JETPLett1973,Gorkov-JETP1974,RiceScott-PRL1975,HirschScalapino-PRL1986}, and which are governed by the same self-consistent mean-field equation, 
\begin{equation}
 \Delta_0=\dosf V\int_{-\omega_0}^{\omega_0}\mathrm{d}\varepsilon_{\vect{k}}\,\frac{[1-2n_\mathrm{F}(E_{\vect{k}})]\Delta_0}{2E_{\vect{k}}}\mathcomma
 \label{eq:gap-equation}
\end{equation}
which determines the order parameter~$\Delta_0$. Here, we abbreviated $E_{\vect{k}}=\sqrt{\varepsilon_{\vect{k}}^2+\Delta_0^2}$, with the energy dispersion~$\varepsilon_{\vect{k}}$. Furthermore, $\dosf V$ is the dimensionless interaction leading to long-range order, $n_\mathrm{F}(E_{\vect{k}})$~is the Fermi distribution, and $\omega_0$ refers to the energetic cut-off of the theory. 
The physical interpretation of~$\Delta_0$ is the pairing gap in the case of superconductivity. In the case of charge-density wave~(CDW) or spin-density wave~(SDW)~ordering, it corresponds to the amplitude of the modulation of the commensurate charge density or spin density, respectively. 
At zero temperature, the gap equation~\eqref{eq:gap-equation} can be integrated straightforwardly, resulting for $\dosf V\ll1$ in the well-known expression for the zero-temperature energy gap 
\begin{equation}
 \Delta_0=2\omega_0\mathrm{e}^{-\frac{1}{\dosf V}}\mathperiod
 \label{eq:gap}
\end{equation}

Experimentally, deviations from mean-field behavior have been observed for density-wave systems in the quantum regime~\cite{JaramilloEtAl-Nature2009,SokolovEtAl-PRB2014,FreitasEtAl-PRB2015,FengEtAl-NPhys2015}. 
The aim of this paper is to theoretically clarify to what extent CDW and SDW~systems fulfill the second characteristic of the BCS~theory, namely a lack of fluctuation corrections as discovered in the case of superconductivity~\cite{GeorgesYedidia-PRB1991,vanDongen-PRL1991,Martin-RoderoFlores-PRB1992,KosMillisLarkin-PRB2004,EberleinMetzner-PRB2013,Eberlein-PRB2014,FischerEtAl2017}. 
We find that the result depends on the order-parameter manifold (in particular, the number of components), and on whether longitudinal or transverse fluctuations, i.\,e., fluctuations of the amplitude or of the phase of the order parameter, dominate. 
In Fig.~\ref{fig:visu-fluctuations}, the different types of fluctuating modes are visualized for three examples. 
In all cases, we show that the characteristic length scales of these fluctuations are within the coherence volume~$\xi^d$, with coherence length~$\xi\sim v_\mathrm{F}/\Delta_0$.

Our analysis focuses on the stability of the ordered state at~$T=0$. A complementary view is the investigation of the instability of the disordered state. For the density-wave systems discussed here, the latter can be performed using a renormalization group~(RG) treatment~\cite{Shankar-RevModPhys1994}. Consistent with our results, channel interference of the RG~analysis suggests corrections to mean-field behavior. This is discussed briefly in Sec.~\ref{sec:rg} and Appendix~\ref{app:two-band}.  
The appeal of our approach is, however, that it offers a simple physical picture for the role of order-parameter fluctuations: amplitude variations suppress the order parameter compared to its mean-field value, whereas phase fluctuations enhance it. 

In this paper, we address the role of fluctuations for mean-field theories of commensurate density-wave order in systems with perfect nesting: We provide a self-consistent calculation of Gaussian fluctuation corrections to the zero-temperature gap equation. We come to the conclusion that, in contrast to BCS~theory, these mean-field theories are not justified since sizable fluctuation corrections are inherent to these theories. 
The mean-field approach is then only valid if the number of amplitude and phase fluctuations is the same. Thus the mean-field approach is neither valid for Heisenberg~SDWs nor for CDWs. 
The only exception is spin-density wave order of XY~spins, where longitudinal and transverse fluctuation contributions cancel exactly -- in accordance with the cancellation of amplitude and phase fluctuations in superconductors~\cite{KosMillisLarkin-PRB2004,FischerEtAl2017}. 
\section{Mean-field theory for density-wave instabilities} 
Density-wave order arises naturally for systems in which the Fermi surface is perfectly nested, i.\,e., different parts of the Fermi surface are connected by the vector~$\vect{Q}$ such that $\varepsilon_{\vect{k}+\vect{Q}}=-\varepsilon_{\vect{k}}$ holds for the energy dispersion. This nesting condition suggests we formulate mean-field theories for density-wave order in complete structural analogy to the BCS~theory of superconductivity. The noninteracting part of the Hamiltonian~$\mathcal{H}=\mathcal{H}_0+\mathcal{H}_\mathrm{int}$ is of the usual form 
\begin{equation} \label{eq:H0}
 \mathcal{H}_0= \sum_{\vect{k},\sigma}\varepsilon_{\vect{k}}\psi^\dagger_{\vect{k},\sigma}\psi^{}_{\vect{k},\sigma}\mathcomma
\end{equation}
where~$\psi_{\vect{k},\sigma}^{(\dagger)}$ annihilates (creates) a fermionic state with momentum~$\vect{k}$ and spin~$\sigma$, where the nesting condition for the dispersion will be implied in the remainder. 
The most general $\mathrm{SU}(2)$-invariant interaction can be written as 
\begin{align}
 \mathcal{H}_\mathrm{int}&= \frac{1}{2}\sum\psi^\dagger_{\vect{k}_1,\sigma_1}\psi^\dagger_{\vect{k}_2,\sigma_2} U_{\sigma_1\sigma_2;\sigma_3\sigma_4}(\vect{k}_1,\vect{k}_2;\vect{k}_3,\vect{k}_4) \nonumber \\ &\qquad \times  \psi^{}_{\vect{k}_3,\sigma_3}\psi^{}_{\vect{k}_4,\sigma_4}\delta(\vect{k}_1+\vect{k}_2-\vect{k}_3-\vect{k}_4)\mathcomma \label{eq:H_interaction}
\end{align}
where the summation is over momenta~$\vect{k}_i$ and spins~$\sigma_i$. 
The spin sector of the interaction can be decomposed into charge~(ch) and spin~(sp) channel according to
\begin{equation}
 U_{\sigma_1\sigma_2;\sigma_3\sigma_4} = U_\mathrm{ch}\,\delta_{\sigma_1\sigma_4}\delta_{\sigma_2\sigma_3}  +U_\mathrm{sp}\,\vect{\sigma}_{\sigma_1\sigma_4}\cdot\vect{\sigma}_{\sigma_2\sigma_3}\mathcomma\label{eq:decomposition}
\end{equation}
where we introduced $\vect{\sigma}=(\sigma_1,\sigma_2,\sigma_3)$ as the vector of Pauli matrices in spin space. 
Furthermore, we could allow for spin-orbit interaction and consider Ising spins or XY~spins instead of the Heisenberg spins introduced in Eq.~\eqref{eq:decomposition} by restricting ourselves to $\vect{\sigma}=(\sigma_1)$ or $\vect{\sigma}=(\sigma_1,\sigma_2)$, respectively. 

When formulating the mean-field theories for charge-density wave~(CDW) order and spin-density wave~(SDW) order, we consider these two channels separately. The interaction projected onto the respective channels takes the form 
\begin{subequations}\label{eq:interaction}
\begin{align}
 \mathcal{H}_\mathrm{ch}&=-\frac{V_\mathrm{ch}}{8}\sum_{\sigma\sigma^\prime}\sum_{\vect{k}\vect{k}^\prime\vect{q}} (\psi^\dagger_{\vect{k},\sigma}\psi^{}_{\vect{k}+\vect{q},\sigma})(\psi^\dagger_{\vect{k}^\prime,\sigma^\prime}\psi^{}_{\vect{k}^\prime-\vect{q},\sigma^\prime}) \mathcomma\\ 
 \mathcal{H}_\mathrm{sp}&=-\frac{V_\mathrm{sp}}{8}\sum_{\sigma_1^{}\sigma_2^{}}\sum_{\sigma_1^\prime\sigma_2^\prime}\sum_{\vect{k}\vect{k}^\prime\vect{q}} (\psi^\dagger_{\vect{k},\sigma_1}\vect{\sigma}_{\sigma_1\sigma_2}\psi^{}_{\vect{k}+\vect{q},\sigma_2}) \nonumber \\ 
 &\qquad \qquad  \cdot (\psi^\dagger_{\vect{k}^\prime,\sigma^\prime_1}\vect{\sigma}_{\sigma_1^\prime\sigma_2^\prime}\psi^{}_{\vect{k}^\prime-\vect{q},\sigma^\prime_2})\mathcomma 
\end{align}
\end{subequations}
where we assumed the couplings $U_{\mathrm{ch},\mathrm{sp}}$ to be independent of momenta and thus introduced $U_{\mathrm{ch},\mathrm{sp}}(\vect{k}_1,\vect{k}_2;\vect{k}_3,\vect{k}_4)\equiv -V_{\mathrm{ch},\mathrm{sp}}/4$. 

The corresponding mean-field theories can be derived from the microscopic Hamiltonian by the usual procedure of introducing an effective bosonic field~$\vect{\Phi}$ via a Hubbard-Stratonovich decoupling
\begin{align}
 &\mathrm{e}^{\frac{V}{2}\sum_{\vect{q}}\vect{b}_{\vect{q}}\cdot\vect{b}_{-\vect{q}}}= \nonumber  \\ 
 &\quad \int\mathcal{D}\vect{\Phi}\,\mathrm{e}^{-\frac{1}{2V}\sum_{\vect{q}}\vect{\Phi}_{\vect{q}}\cdot\vect{\Phi}_{-\vect{q}}+\frac{1}{2}\sum_{\vect{q}}\vect{\Phi}_{\vect{q}}\cdot\vect{b}_{-\vect{q}}+\frac{1}{2}\sum_{\vect{q}}\vect{b}_{\vect{q}}\cdot\vect{\Phi}_{-\vect{q}}}\label{eq:decoupling}
\end{align}
 of the interaction~\eqref{eq:interaction} in the channel of interest and subsequently integrating out the fermions. This requires~$V>0$ for the interaction in the respective channel, where we skip the subindex ch or sp in what follows. The effective field~$\vect{\Phi}\in\mathds{R}^N$ plays the role of an order parameter and its dimensionality~$N$ depends on the channel in which the decoupling of the interaction is performed. 
Formulated in the language of field integrals, the corresponding mean-field theory follows immediately at the level of the saddle-point approximation, i.\,e., assuming that the order parameter be temporally homogeneous and $\delta$-distributed in momentum space, which we denote~$\vect{\Phi}_0$. 

The mean-field theories for density-wave order resulting from nesting bear the same structure as the BCS~theory of superconductivity. 
The ground-state energy 
\begin{equation}
 E_\mathrm{MF}(\Phi_0)=E_0-2\dosf L^d\Phi_0^2\bigg[\ln\bigg(\frac{2E_\mathrm{F}}{\Phi_0}\bigg)+\frac{1}{2}-\frac{1}{\dosf V}\bigg] 
\end{equation}
deep inside the ordered phase (where $\Phi_0\equiv|\vect{\Phi}_0|$ is the magnitude of the order parameter) is reduced as compared to the high-temperature ground-state energy~$E_0$. 
For the instantaneous electronic interaction~\eqref{eq:interaction} considered here, the energy window of the attractive pairing goes up to the Fermi energy. That means, in contrast to the BCS~theory where the energetic cut-off $\omega_0$ of the theory is given by the Debye energy, we consider $\omega_0\simeq E_\mathrm{F}$ in the remainder. 
Furthermore, $\dosf$~refers to the density of states at the Fermi level and $L^d$ denotes the system's volume. 
This structure results in the characteristic logarithm appearing in the zero-temperature mean-field gap equation 
\begin{equation}
0=\frac{1}{\dosf L^d}\frac{\dd E_\mathrm{MF}(\Phi_0)}{\dd\Phi_0^2} 
\approx \frac{1}{\dosf V}-\ln\bigg(\frac{2E_\mathrm{F}}{\Phi_0}\bigg)\mathcomma
\label{eq:gap-equation2}
\end{equation}
which self-consistently determines the magnitude 
of the mean-field value of the order parameter. 
Details on the derivation can be found in Appendix~\ref{app:calculation}. 
Note that Eq.~\eqref{eq:gap-equation2} is equivalent to Eq.~\eqref{eq:gap-equation} at $T=0$. 
The logarithmic contribution arises from a relative sign provided by the nesting condition~$\varepsilon_{\vect{k}+\vect{Q}}=-\varepsilon_{\vect{k}}$ in the case of density-wave order, whereas for superconductivity it results from the pairing of time-reversed states. 
Hence, the density-wave analog for the Nambu spinors is given by 
\begin{equation}
 \Psi_{\vect{k}}=\begin{pmatrix} \psi^\dagger_{\vect{k},\uparrow} & \psi^\dagger_{\vect{k},\downarrow} & \psi^\dagger_{\vect{k}+\vect{Q},\uparrow} &\psi^\dagger_{\vect{k}+\vect{Q},\downarrow}\end{pmatrix}^T \mathcomma \label{eq:spinor}
\end{equation}
allowing us to represent the mean-field ground-state energy as $E_\mathrm{MF}(\Phi_0)=2\Phi_0^2/V-\int_k\tr\ln[-(\mathcal{G}_k^\mathrm{MF})^{-1}]$. The corresponding matrix Green's function in Matsubara representation is given by 
\begin{align}
 &\big(\mathcal{G}_k^{\mathrm{MF}}\big)^{-1}=\frac{1}{2}\left(\mathrm{i}\nu_n\tau_0\sigma_0-\varepsilon_{\vect{k}}\tau_3\sigma_0\right) \nonumber \\ 
  &\qquad +\frac{1}{2}\tau_1\left\{\begin{matrix}\Phi_0\sigma_0 & \text{for CDW order} \mathcomma\\ \vect{\Phi}_0\cdot \vect{\sigma}& \text{for Heisenberg SDW order}\mathcomma\end{matrix}\right.
\end{align}
where we abbreviated $k\equiv(\vect{k},\nu_n)$ with the fermionic Matsubara frequency $\nu_n=(2n+1)\pi T$. The Pauli matrices~$\sigma_i$ refer to the spin sector whereas the $\tau_i$ denote Pauli matrices in the band space that emerges due to a doubling of the unit cell. We assume~$2\vect{Q}$ to be a reciprocal lattice vector.
\section{Fluctuation corrections to the gap equation} 
The validity of the mean-field approximation as discussed in the previous section relies on the assumption that deviations of the order parameter from its mean-field value lead to negligible corrections to the mean-field theory in the sense that fluctuation corrections to physical observables are small compared to their mean-field value. 

Of course, there are regimes in which the role of long-wavelength ($|\vect{q}|\ll2\xi^{-1}$) fluctuations is well-understood. Firstly, thermal fluctuations drive the phase transition. The Ginzburg regime in which these thermal fluctuations lead to sizable corrections to mean-field theory is restricted to the vicinity of the critical point. 
Secondly, in low-dimensional systems fluctuations are important down to lower temperatures, where they can lead to the breakdown of true long-range order~\cite{MerminWagner-PRL1966,Hohenberg-PR1967}.
At zero temperature, deep inside the ordered phase, (quantum) fluctuations are expected to have severe consequences only for one-dimensional systems, while they are small in spatial dimensions~$d\geq2$. 
Here we consider different fluctuations with characteristic length scales shorter than~$\xi$. 

To self-consistently check the validity of a mean-field theory, we determine fluctuation corrections to the zero-temperature gap equation and compare the resulting contributions due to fluctuations to the mean-field terms. This approach was put forward in the context of superconductivity by Kos, Millis, and Larkin~\cite{KosMillisLarkin-PRB2004}; and their self-consistent calculation of corrections to the BCS~gap equation indeed showed that the BCS~mean-field theory is justified. This analysis for $s$-wave superconductors can be extended to charged superconductors~\cite{FischerEtAl2017} as well as anisotropic superconductors~\cite{ParamekantiEtAl-PRB2000,BarlasVarma-PRB2013,Hoyer-Thesis2017}, resulting in the conclusion that fluctuation corrections to the zero-temperature gap equation are generally negligible for superconductors. Remarkably, this is due to an exact cancellation of individually large contributions that can be assigned to fluctuations of the amplitude and the phase of the order parameter.  Therefore, a natural question in the context of density-wave instabilities is whether an analogous mechanism of cancellation of longitudinal and transverse fluctuation corrections ensures the validity of the respective mean-field theories -- or whether quantum fluctuations become sizable such that the mean-field description is not justified. This question will be addressed in the remainder of this paper. 

For density-wave instabilities, the order parameter governing the effective action is a real $N$-component vector, $\boldsymbol{\Phi}(\boldsymbol{r},\tau)\in\mathds{R}^N$, where~$\tau$ refers to the  imaginary time in Matsubara formalism. This order parameter can be split into the static and homogeneous mean-field value and spatial and temporal fluctuations around this mean-field value as
\begin{equation}
 \boldsymbol{\Phi}(\boldsymbol{r},\tau)=\Phi_0\hat{\boldsymbol{e}}+\delta\boldsymbol{\Phi}(\boldsymbol{r},\tau)\mathperiod
\end{equation}
Here, we introduced~$\hat{\boldsymbol{e}}$ as the unit vector along the direction of the mean-field order parameter and the magnitude of the mean-field order parameter~$\Phi_0$ can be determined self-consistently from the gap equation. The fluctuations~$\delta\boldsymbol{\Phi}$ of the order parameter around the mean-field configuration can be further split into one longitudinal mode~$\parallel \hat{\boldsymbol{e}}$ and $N-1$~transverse modes~$\perp \hat{\boldsymbol{e}}$. In the remainder of this paper, we restrict ourselves to the leading contribution and evaluate the fluctuation corrections for Gaussian fluctuations, i.\,e., taking into account contributions up to~$\mathcal{O}[(\delta\vect{\Phi})^2]$. 

\begin{figure}[t]
\includegraphics[width=\columnwidth]{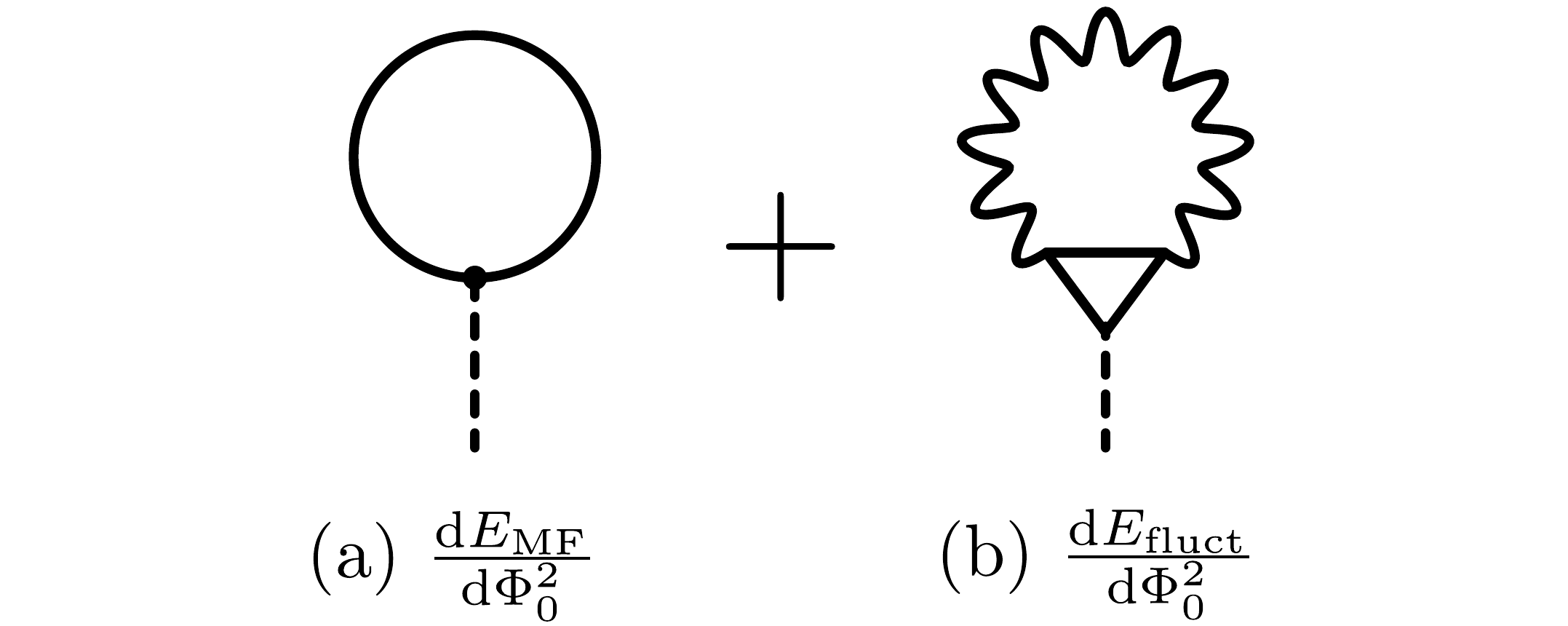}\centering
\caption{Diagrammatic representation of contributions to the gap equation. Straight lines represent the fermionic propagators while wiggly lines stand for the fluctuation propagator. The order parameter (indicated by a dashed line) is added for the sake of clarity here, however, it does not contribute to the derivative~$\dd E/\dd\Phi_0^2$. The logarithmic contribution to the mean-field gap equation is shown in~(a), while the structure of Gaussian fluctuation corrections to the gap equation is presented in~(b).}
\label{fig:gap-equation}
\end{figure}
\begin{figure}[t]
 \includegraphics[width=\columnwidth]{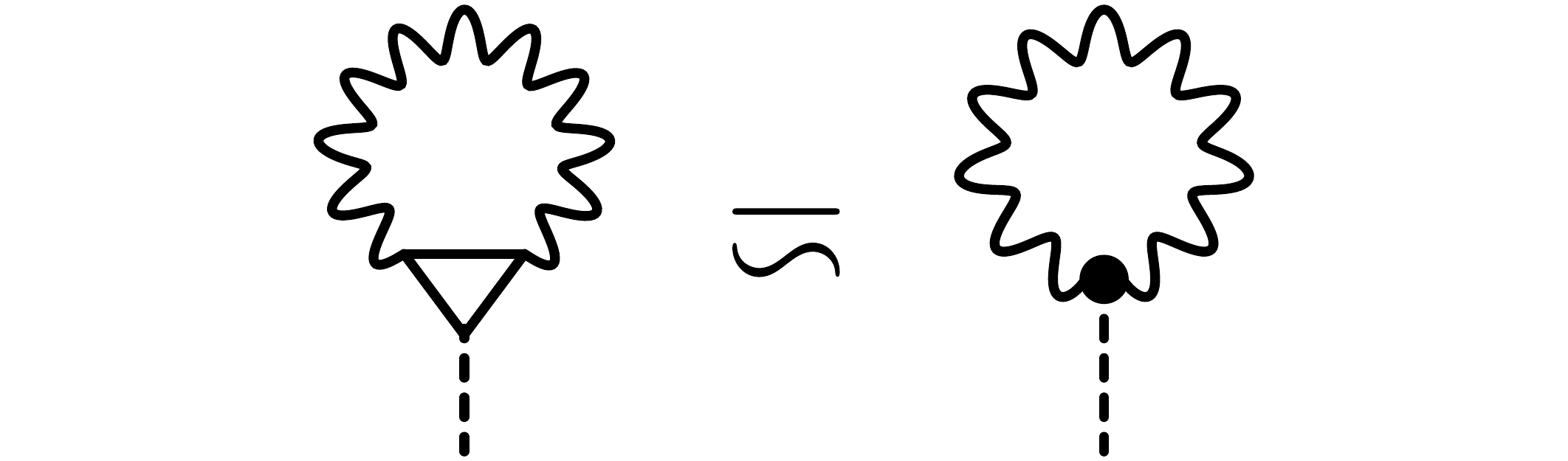}\centering
 \caption{Fluctuation corrections to the gap equation in the long-wavelength regime~($|\vect{q}|\ll2\xi^{-1}$) and for small frequencies~($\omega\ll2\Phi_0$). The fermionic triangle part of the fluctuation corrections to the gap equation [also shown in Fig.~\ref{fig:gap-equation}(b)] becomes structureless. This limit corresponds to the usual contributions known from the consideration of critical fluctuations.  Note that in the opposite regime of short-wavelength fluctuations ($|\vect{q}|\gg2\xi^{-1}$), the inner structure of the fermionic triangle becomes crucial, cf.\ Fig.~\ref{fig:triangles}.}
 \label{fig:long-wavelength}
\end{figure}

Including fluctuations around the saddle-point configuration results in additional contributions to the ground-state energy, $E(\Phi_0)=E_\mathrm{MF}(\Phi_0)+E_\mathrm{fluct}(\Phi_0)$, which are also reflected in the gap equation as 
\begin{equation}
\frac{1}{\dosf L^d}\frac{\dd E(\Phi_0)}{\dd \Phi_0^2}=\frac{1}{\lambda}-\ln\Big(\frac{2E_\mathrm{F}}{\Phi_0}\Big)+\frac{1}{\dosf L^d}\frac{\dd E_\mathrm{fluct}(\Phi_0)}{\dd\Phi_0^2}=0\mathcomma \label{eq:gap-equation-fluct}
\end{equation}
where we introduced the dimensionless interaction~$\lambda=\dosf V$. 
Unless the additional contribution stemming from fluctuations is smaller than the first two terms already appearing at mean-field level [cf.~Eq.~\eqref{eq:gap-equation2}], the mean-field theory is not justified. 
The logarithmic contribution already appearing at mean-field level corresponds to the fermionic loop diagram depicted in Fig.~\ref{fig:gap-equation}(a). 
The Gaussian fluctuation corrections studied here take the form presented in Fig.~\ref{fig:gap-equation}(b). In the remainder, we show that such terms indeed give rise to a logarithmic contribution to the gap equation. 
The usual contributions due to critical fluctuations, important in the long-wavelength limit and for small frequencies, correspond to the limit where the fermionic triangle becomes structureless, cf.~Fig.~\ref{fig:long-wavelength}, but our analysis shows that the internal structure of the triangle diagram is indeed important. 

If the Gaussian fluctuation corrections are of the same order as the mean-field contribution to the zero-temperature gap equation, they can be effectively restated as a modification of the prefactor of the logarithmic contribution. Then the solution of Eq.~\eqref{eq:gap-equation-fluct} can be written in the familiar form~\eqref{eq:gap} 
by introducing the effective interaction 
\begin{equation} 
 \lambda_\mathrm{eff}=\lambda\bigg[1-\frac{1}{\dosf L^{d}\ln\big(\frac{2 E_\mathrm{F}}{\Phi_0}\big)}\frac{\dd E_\mathrm{fluct}(\Phi_0)}{\dd\Phi_0^2}\bigg]\mathperiod
\end{equation}
Hence the sign of the fluctuation contribution plays an interesting role as negative fluctuation corrections further enhance the gap as compared to its mean-field value whereas positive fluctuation contributions result in a reduction, i.\,e., weaken the ordered state. 

Let us now analyze the structure of fluctuation corrections in the case of density-wave instabilities due to nesting following the logic of Ref.~\onlinecite{KosMillisLarkin-PRB2004}, also building on the calculations presented in Ref.~\onlinecite{FischerEtAl2017}. 
More details specific to our derivation for both charge-density wave order and spin-density wave order are presented in Appendix~\ref{app:calculation}. 
The Gaussian fluctuation corrections can be expressed in terms of the fluctuation propagator~$\boldsymbol{\mathcal{D}}_q$ as $E_\mathrm{fluct}(\Phi_0)=\frac{L^d}{2}\int_q\ln\det(\vect{\mathcal{D}}^{-1}_q)$, where the dimensionality of the order parameter translates to the dimensionality of the fluctuation propagator. 
Accordingly, the Gaussian fluctuation corrections to the zero-temperature gap equation take the form 
\begin{equation}
 \frac{1}{\dosf L^d}\frac{\dd E_\mathrm{fluct}(\Phi_0)}{\dd\Phi_0^2}=\frac{1}{2\dosf}\int_q\,\frac{\frac{\dd}{\dd\Phi_0^2}\det(\boldsymbol{\mathcal{D}}^{-1}_q)}{\det(\boldsymbol{\mathcal{D}}^{-1}_q)}\mathcomma 
 \label{eq:fluctuation-corrections}
\end{equation}
where we introduced $q\equiv(\vect{q},\omega)$ and the integration $\int_q\ldots=\int\frac{\dd\omega}{2\pi}\int\frac{\dd^d\vect{q}}{(2\pi)^d}\ldots$ runs over all external frequencies and momenta up to the cutoff. 
Consequently, an expansion of the fluctuation spectrum in small momenta and frequencies as often discussed in the context of collective modes is not sufficient here. 
Instead, the fluctuation propagator has to be determined for all frequencies and momenta, and in fact, the short-wavelength fluctuations with momenta~$2\xi^{-1}=2\Delta_0/v_\mathrm{F}\ll|\vect{q}|\ll k_\mathrm{F}$ turn out to be crucial, as they give rise to an additional logarithmic contribution to the zero-temperature gap equation as discussed in Sec.~\ref{sec:results}. 

The inverse matrix of the fluctuation propagator for density-wave order can be stated in terms of the polarization matrix~$\boldsymbol{\Pi}_q$ as  
\begin{equation}
 \boldsymbol{\mathcal{D}}^{-1}_q=\frac{1}{V}\boldsymbol{\mathds{1}}-\boldsymbol{\Pi}_q\mathperiod
\end{equation}
For density-wave order, transverse and longitudinal fluctuations are not coupled and hence the polarization matrix is diagonal. The Gaussian fluctuation corrections can thus be expressed in terms of longitudinal and transverse contributions using 
\begin{equation}
 E_\mathrm{fluct}(\Phi_0)=\frac{L^d}{2}\int_q\ln\Big[4^N\Big(\frac{1}{V}-\Pi^\perp_q\Big)^{N-1}\Big(\frac{1}{V}-\Pi^\parallel_q\Big)\Big]\mathperiod 
\end{equation}
Both the longitudinal and the transverse contribution consist of a normal part~$\Pi^\mathrm{n}_q=\frac{1}{2}\int_k(G_{k+q}G_{-k}+G_{-k}G_{k-q})$ which survives the limit~$\Phi_0\rightarrow0$, and an anomalous contribution~$\Pi^\mathrm{a}_q=-\int_kF_kF_{k+q}$ which vanishes in the high-temperature normal state, 
\begin{align}
 \Pi^\perp_q&=\Pi^\mathrm{n}_q-\Pi^\mathrm{a}_q\mathcomma \label{eq:Pi-perp} \\
 \Pi^\parallel_q&=\Pi^\mathrm{n}_q+\Pi^\mathrm{a}_q\mathperiod \label{eq:Pi-parallel}
\end{align}
These integrals (see Appendix~\ref{app:polarization-function} for details) are of the same structure as those arising in the context of superconductivity, and hence we can build on the results obtained by previous studies~\cite{KosMillisLarkin-PRB2004,FischerEtAl2017} in the remainder.
\section{Results and Discussion}\label{sec:results}
\begin{figure}[t]
\includegraphics[width=\columnwidth]{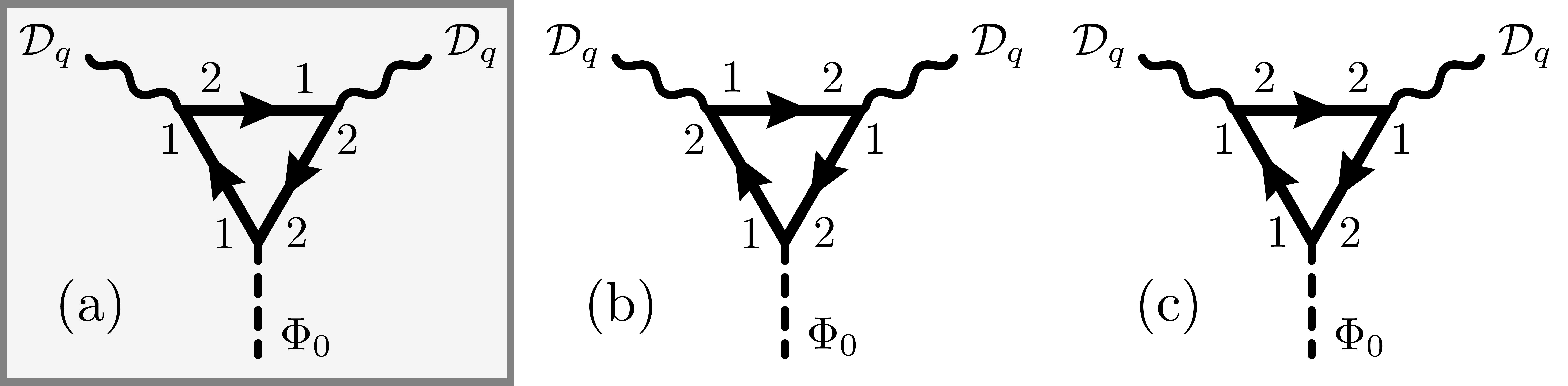}\centering
\caption{Diagrammatic representation of the terms contributing to the derivative of the polarization function. For the definition of the diagrammatic elements see Fig.~\ref{fig:diagrammatic-elements}. (a) and (b) Derivatives of the anomalous part, while (c) represents the derivative of the normal contribution. 
While in the long-wavelength regime, the fermionic triangle part depicted here becomes structureless (cf.\ Fig.~\ref{fig:long-wavelength}), it is of particular importance for the contributions stemming from the regime of short wavelengths. The diagram which yields the crucial contribution to the gap equation discussed in this paper is highlighted in gray.} 
\label{fig:triangles}
\end{figure}
We can straightforwardly adopt the formalism developed in the context of superconductivity~\cite{KosMillisLarkin-PRB2004,FischerEtAl2017} to calculate the leading-order corrections to the zero-temperature mean-field gap equation due to Gaussian fluctuations of the order parameter for density-wave order. 
We consider the regimes of small and large momenta/frequencies separately, which is possible~\cite{VaksGalitskiiLarkin-JETP1962} since the integrals only depend on the combination~$r=\sqrt{\omega^2+(v_\mathrm{F}|\vect{q}|\cos\theta)^2}/(2\Phi_0)$, where~$\theta$ denotes the angle between fermionic momentum~$\vect{k}$ and bosonic momentum~$\vect{q}$. 
The crucial contribution to the fluctuation corrections, as given by Eq.~\eqref{eq:fluctuation-corrections} and diagrammatically represented in Fig.~\ref{fig:gap-equation}(b), stems from the regime where $r\gg1$ and $v_\mathrm{F}|\vect{q}|>\omega$, whereas long-wavelength fluctuations lead to corrections that are negligible compared to the mean-field terms. This is due to the fact that the fermionic triangle (cf.~Fig.~\ref{fig:triangles}) associated with the derivative of the polarization function becomes structureless in the limit~$r\ll1$ and hence fluctuation corrections reduce to the simpler form shown in Fig.~\ref{fig:long-wavelength}. 

In the regime of interest, the fluctuation propagator is dominated by the normal part of the polarization function, while its derivative is largely determined by the anomalous part. Therefore, the leading contribution to the gap equation evaluates to 
\begin{subequations}\label{eq:result}\begin{align}
  &\frac{1}{\dosf L^d}\frac{\dd E_\mathrm{fluct}(\Phi_0)}{\dd\Phi_0^2}\Big|_{r\gg1} \approx \frac{1}{2\dosf}\int_q\frac{[(N-1)-1]\frac{\dd\Pi^\mathrm{a}_q}{\dd \Phi_0^2}}{\frac{1}{V}-\Pi^\mathrm{n}_q} \\ 
  &\qquad \qquad \approx -\frac{1}{2}\frac{(N-1)-1}{8\pi (d-1)}\ln\Big(\frac{E_\mathrm{F}}{\Phi_0}\Big)\mathperiod
\end{align}\end{subequations}
This reveals that Gaussian fluctuations indeed quite generically entail an additional logarithmic contribution to the gap equation for density-wave instabilities due to nesting. 
This logarithmic divergence can be traced back to the fermionic triangle depicted in Fig.~\ref{fig:triangles}(a), which results from the derivative of the anomalous part of the polarization function that enters the numerator of $\dd E_\mathrm{fluct}(\Phi_0)/\dd\Phi_0^2$, while the two other contributions [Figs.~\ref{fig:triangles}(b) and \ref{fig:triangles}(c)] do not give rise to additional logarithms. 
Thus, the precarious process that invalidates the mean-field approach in the case of density-wave order is the interaction of quasiparticles through fluctuations. 

\begin{table}[t]\centering
\begin{tabular}{llr}\hline 
 \multicolumn{2}{l}{type of order}  & \qquad fluctuations\\ \hline 
 Heisenberg~SDW &  \quad($N=3$) & \qquad increase gap \\
 XY~SDW  & \quad ($N=2$) & are negligible \\
 Ising~SDW or CDW  & \quad ($N=1$) & decrease gap  \\ \hline 
\end{tabular}
\caption{Gaussian fluctuation corrections in $\mathrm{O}(N)$~models. Depending on the dimensionality of the order parameter, fluctuation corrections can increase or decrease the gap value compared to its mean-field value.} 
\label{tab:summary-sdw-fluctuations}
\end{table}
The above result is in stark contrast to the insignificant role that fluctuations play in the context of superconductivity: Analogous contributions to the BCS~gap equation are not only suppressed by the smallness of the Debye energy as compared to the Fermi energy, but the corresponding contributions stemming from fluctuations of phase and amplitude of the complex order parameter even cancel exactly. 
Nonetheless, the corrections due to the longitudinal mode and the $N-1$~transverse modes enter the result~\eqref{eq:result} with opposite signs, and hence the prefactor depends on the number of transverse modes. 
Only for XY~spins ($N=2$), the large fluctuation corrections stemming from the regime~$r\gg1$ cancel -- which is consistent with previous results in the context of superconductivity~\cite{KosMillisLarkin-PRB2004,FischerEtAl2017}. 
In conclusion, the analysis of the role of fluctuation corrections in the case of density-wave order reveals that mean-field approaches are generally not justified in this situation, the only exception being spin-density wave order of XY~spins. 

Furthermore, the sign of the fluctuation corrections allows us to judge whether the presence of fluctuations is advantageous or detrimental to the formation of density-wave order: The effective interaction 
\begin{align}
 \lambda_\mathrm{eff}
 &= \lambda\bigg[1+\frac{N-2}{16\pi(d-1)}\bigg]\mathcomma 
 \label{eq:effective-interaction}
\end{align}
which governs the ordering in the presence of fluctuations, either decreases ($\dd E_\mathrm{fluct}/\dd \Phi_0^2>0$) or increases ($\dd E_\mathrm{fluct}/\dd\Phi_0^2<0$), and the same is true for the solution of the zero-temperature gap equation~\eqref{eq:gap-equation2} in the presence of fluctuations. 
For $N=1$ or $N=3$ and $d=3$, the relative change in~$\lambda$ is $1/(32\pi)\simeq0.01$, i.\,e., rather small. More important than the numerical value of this correction is the fact that there is no guarantee that even higher-order processes give rise to equally non-negligible corrections. 
Our analysis of fluctuation corrections shows that contributions stemming from the longitudinal mode lead to a decrease of the gap compared to its mean-field value, whereas transverse fluctuations increase the gap. Hence, if the latter dominate, fluctuations are favorable to the formation of density-wave order, as it is the case for spin-density wave order of Heisenberg spins. If, on the other hand, transverse fluctuations cannot compensate for the effect of the longitudinal mode, the ordered state is weakened. This applies to mean-field theories for charge-density wave order as well as for spin-density wave order of Ising spins, see also Table~\ref{tab:summary-sdw-fluctuations} for an overview of our results. 
\section{Channel interference in the renormalization group approach}\label{sec:rg}
\begin{figure*}[t]
 \includegraphics[width=\textwidth]{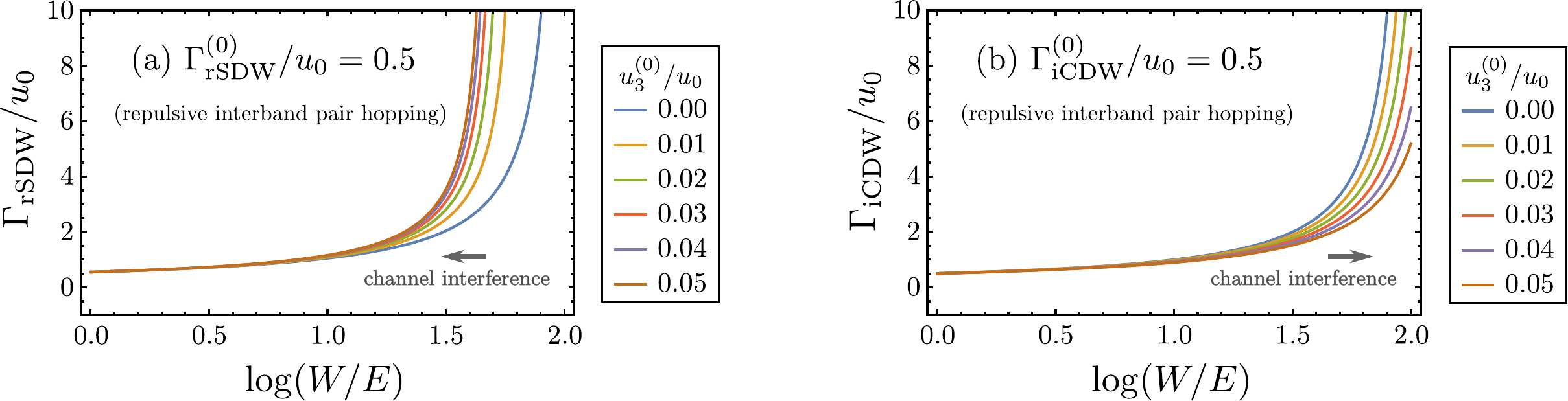}
 \caption{Effect of channel interference on density-wave instabilities resulting from repulsive interband pair hopping, see Appendix~\ref{app:two-band} for more details. (a) Flow of $\Gamma_\mathrm{rSDW}$ for fixed bare interaction~$\Gamma_\mathrm{rSDW}^{(0)}=0.5$ while $\Gamma_\mathrm{iCDW}^{(0)}=0$ and $\Gamma_{s^{+-}}^{(0)}=0$: Increasing $u_3^{(0)}$ (and hence channel-interference strength) results in higher transition temperatures. (b) Flow of $\Gamma_\mathrm{iCDW}$ for fixed bare interaction~$\Gamma_\mathrm{iCDW}^{(0)}=0.5$ while $\Gamma_\mathrm{rSDW}^{(0)}=0$ and $\Gamma_{s^{+-}}^{(0)}=0$: Here, increasing $u_3^{(0)}$ implies lower transition temperatures.}
 \label{fig:channel-interference}
\end{figure*}
Our analysis of the role of fluctuations for density-wave instabilities has been performed at zero temperature, i.\,e., deep inside the ordered phase. 
In doing so, we concentrated on a single channel and neglected the potential presence of competing instabilities. 
Another, complementary perspective is provided by a renormalization group~(RG) analysis~\cite{Shankar-RevModPhys1994}, in which competing instabilities of the system can be treated on equal footing and thereby channel interference can be studied within this framework. However, within the RG~scheme, the phase transition is approached coming from high energies, allowing us to determine the leading instability of the system, but this approach is less suited to explore the ordered state further. For example, the model defined by the Hamiltonian~$\mathcal{H}=\mathcal{H}_0+\mathcal{H}_\mathrm{int}$ as stated in Eqs.~\eqref{eq:H0} and~\eqref{eq:H_interaction} is in principle also unstable towards the formation of superconductivity, which is expected to interfere with SDW~ordering. 

In fact, the complex phase diagrams of materials of current interest such as the iron-based superconductors can be understood as a result of the interplay of different competing instabilities: Already the simple two-band model allows for superconductivity as well as SDW order and CDW order~\cite{ChubukovEfremovEremin-PRB2008,PodolskyKeeKim-EPL2009,KangTesanovic-PRB2011}, owing to the nested nature of the Fermi surface. 
This model is essentially a band-basis translation of the model investigated in this paper. Hence the calculation of fluctuation corrections to the corresponding zero-temperature gap equations is straightforward, see Appendix~\ref{app:calculation}. 
What is more, we can use the RG~equations derived for the same model~\cite{ChubukovEfremovEremin-PRB2008} to analyze the effect of channel interference on the transition temperature towards a given ordered state. 
The resulting flow of the couplings in the density-wave channels as a function of $t=\log\frac{W}{E}$, where $W$ is the bandwidth and $E$ the running energy scale, takes the form 
\begin{subequations}\label{eq:channel-interference-couplings}
 \begin{align}
  \dot{\Gamma}_\mathrm{SDW}&= (\Gamma_\mathrm{SDW})^2\pm 2u_3(u_1-u_2-u_4)\mathcomma \\ 
  \dot{\Gamma}_\mathrm{CDW}&= (\Gamma_\mathrm{CDW})^2\mp 2u_3(u_1+u_2-u_4) \mathperiod
 \end{align}
 \end{subequations}
The couplings $u_i$ refer to different intraband and interband processes that are connected to the couplings in spin and charge channel according to Eq.~\eqref{eq:couplings-connection}. 
If the second term were zero, this would result in the usual logarithmic divergence~$\Gamma=\Gamma_0/(1-\Gamma_0\log\frac{W}{E})$ of the coupling in the respective channel. Hence the presence of the second term implies corrections due to channel interference that are intrinsic to the model as long as the interband pair-hopping process which is associated with~$u_3$, and involves a momentum transfer of $2\vect{Q}$, is effective. Motivated by the structure of Eq.~\eqref{eq:channel-interference-couplings}, we used~$u_3^{(0)}$ as a measure of channel interference in our brief analysis, where we tuned the bare couplings such that the instability in the channel under consideration is favored within the mean-field description, while the bare couplings in the competing channels were tuned to zero. We then kept the bare value of the coupling in a given channel fixed while increasing the channel interference strength. 
Details on our derivation can be found in Appendix~\ref{app:two-band}. We find that the energy scale at which the coupling diverges is affected by channel interference. Furthermore, the effect of channel interference on charge-density wave order and spin-density wave order is different: While increasing the channel-interference strength via~$u_3^{(0)}$ results in higher transition temperatures for SDW~order (since the divergence is pushed to lower energies), the opposite is true for CDW~order. In Fig.~\ref{fig:channel-interference}, we present our numerical solution of the RG~equations~\eqref{eq:channel-interference-couplings} for SDW~order characterized by a real order parameter and CDW~order associated with a purely imaginary order parameter, since these are the two instabilities arising from the parameter range usually discussed in the context of iron-based superconductors. 

While our analysis and the perturbative RG~investigation address rather different phenomena, we note that the trends we find from the RG~flow are consistent with our analysis of fluctuation corrections:  Channel interference is favorable for the formation of spin-density wave order as it leads to an increase of the corresponding transition temperature, whereas the transition towards charge-density wave order is hindered by channel interference in the sense that the corresponding transition temperature decreases. 

\section{Summary} 
Mean-field theories for density-wave order resulting from a nesting of the Fermi surface can be derived from a microscopic model of interacting fermions in full analogy to the formulation of BCS~theory in the language of field integrals. 
The BCS~theory of superconductivity is the poster child of mean-field theories since neither thermal nor quantum fluctuations lead to sizable effects in conventional superconductors and hence can be neglected. 
In this paper, we showed that, in contrast to superconductivity (where the order parameter is a complex scalar), the impact of fluctuations is crucial in the case of commensurate density-wave order (characterized by a real $N$-component order parameter) as long as $N\neq2$. 

To be specific, we have investigated the role of fluctuations for charge-density wave order and spin-density wave order due to nesting of the Fermi surface. 
Our main finding is that, generally,  fluctuation corrections to the zero-temperature gap equations for such density-wave instabilities are of the same order~$\mathcal{O}[\ln(E_\mathrm{F}/\Phi_0)]$ as the terms already appearing at mean-field level. 
In conclusion, the mean-field theories for density-wave instabilities are not justified since the large fluctuation corrections imply that the respective mean-field theory cannot capture all relevant contributions. 
Of course, our analysis does not imply that the effect of fluctuations is merely to replace~$\lambda$ by $\lambda_\mathrm{eff}$ of Eq.~\eqref{eq:effective-interaction}, as there is no guarantee that even higher-order processes will not give rise to corrections of the same order. 
Moreover, we find that the additional logarithmic contribution to the gap equation stemming from fluctuations originates from the derivative of the anomalous polarization function, and the crucial process is the interaction of quasiparticles through the exchange of fluctuations depicted in Fig.~\ref{fig:triangles}(a). 

Interestingly, we find that the impact of longitudinal and transverse modes is quite different: Longitudinal fluctuations always yield $E_\mathrm{fluct}>0$ and are thus detrimental to the formation of long-range order. Transverse fluctuations, on the other hand, only yield $E_\mathrm{fluct}>0$ in the long-wavelength regime. In the opposite regime of transverse fluctuations on lengthscales smaller than the coherence length, we find that the respective fluctuation correction surprisingly lowers the energy. Furthermore, this contribution is the dominant one since it yields the additional logarithm to the gap equation in the case of perfect nesting that ultimately leads to an increase of the gap as compared to its mean-field value.
It is due to the twist of the ``phase'' induced by the excitation of quasiparticles inside the coherence volume which enhances the kinetic energy of quasiparticle excitations. In contrast, longitudinal fluctuations can only lead to an increase of the energy since the potential energy is already minimized by the mean-field configuration.

Because of the different nature of longitudinal and transverse fluctuations, the case~$N=2$ is an interesting exception: The logarithmic contributions to the gap equation stemming from the longitudinal mode and the single transverse mode cancel exactly which renders the overall fluctuation corrections negligible. This cancellation legitimates the mean-field approach to density-wave order of XY~spins. This is in accordance with the analysis of fluctuations in the context of superconductivity~\cite{KosMillisLarkin-PRB2004,FischerEtAl2017}, where fluctuation corrections from phase and amplitude mode cancel analogously, providing the justification of the BCS~mean-field theory. 
\section{Acknowledgments}
We thank M.~Bard, A.~V.~Chubukov, S.~Fischer, M.~Hecker, N. Kainaris, and M.~S.~Scheurer for helpful discussions.
\appendix
\section{Calculation of fluctuation corrections}\label{app:calculation}
This Appendix provides technical details on our calculation of fluctuation corrections, closely following the logic and notation introduced in the context of superconductivity in Ref.~\onlinecite{KosMillisLarkin-PRB2004} and extended by Ref.~\onlinecite{FischerEtAl2017}. 

In what follows, we consider the partition function $\mathcal{Z}=\int\mathcal{D}[\bar{\psi},\psi]\,\exp(-\mathcal{S}[\bar{\psi},\psi])$ with the appropriate action $\mathcal{S}[\bar{\psi},\psi]=\int_0^\beta\dd\tau\,(\sum_\sigma\int\dd^d\vect{x}\,\bar{\psi}_\sigma(\vect{x})\partial_\tau\psi_\sigma(\vect{x})+\mathcal{H})$ corresponding to the Hamiltonian as stated in Eqs.~\eqref{eq:H0} and~\eqref{eq:H_interaction}. The effective theory in terms of the order parameter~$\vect{\Phi}$ follows from the Hubbard-Stratonovich transformation~\eqref{eq:decoupling} and successively integrating out the fermions, resulting in $\mathcal{Z}=\int\mathcal{D}\vect{\Phi}\,\exp(-\mathcal{S}_\mathrm{eff}[\vect{\Phi}])$. 

Since the derivation of fluctuation corrections to the zero-temperature mean-field gap equation for charge-density wave~(CDW) order and spin-density wave~(SDW) order follow the same logic and primarily differ in the dimensionality of the respective order parameter~$\vect{\Phi}\in\mathds{R}^N$, we treat them simultaneously here. 
For CDW~order, the order parameter (associated with the charge density~$\rho$) introduced by the Hubbard-Stratonovich transformation~\eqref{eq:decoupling} is a scalar, 
\begin{equation}
\rho_{\vect{q}}=\sum_{\vect{k},\sigma,\sigma^\prime}\big<\psi_{\vect{k},\sigma}^\dagger\delta_{\sigma\sigma^\prime}\psi_{\vect{k}+\vect{Q}+\vect{q},\sigma^\prime}^{}\pm\psi_{\vect{k}+\vect{Q},\sigma}^\dagger\delta_{\sigma\sigma^\prime}\psi_{\vect{k}+\vect{q},\sigma^\prime}^{}\big>\mathcomma 
\end{equation}
corresponding to $N=1$. The upper sign refers to CDW~order characterized by a real~(r) order parameter, whereas the lower sign refers to CDW~order with an imaginary~(i) order parameter. The latter follows from assuming $V<0$ in the respective channel. All derivations for iCDW~order can be performed in complete analogy to those for rCDW~order and since we come to the same conclusions in both cases, we concentrate on rCDW order in the following. 
For SDW~order of Heisenberg spins, the order parameter (associated with the magnetization~$\vect{M}$) is a three-component vectorial object, 
\begin{equation}
\vect{M}_{\vect{q}}=\sum_{\vect{k},\sigma,\sigma^\prime}\big<\psi_{\vect{k},\sigma}^\dagger\vect{\sigma}_{\sigma\sigma^\prime}\psi_{\vect{k}+\vect{Q}+\vect{q},\sigma^\prime}^{}\pm\psi_{\vect{k}+\vect{Q},\sigma}^\dagger\vect{\sigma}_{\sigma\sigma^\prime}\psi_{\vect{k}+\vect{q},\sigma^\prime}^{}\big>\mathcomma 
\end{equation}
which corresponds to~$N=3$. Again, for the sake of clarity, we only discuss rSDW~order (upper sign) since, mutatis mutandis, the same results can be obtained for iSDW~order (lower sign). 

For both types of density-wave order, the order parameter (which we denote~$\vect{\Phi}$ henceforth) can be split into the static and homogeneous mean-field value~$\Phi_0$ and fluctuations~$\delta\vect{\Phi}$ around this mean-field value as
\begin{equation}
 \vect{\Phi}_q=\Phi_0\hat{\vect{e}}\delta_{q,0}+\delta\vect{\Phi}_q\mathcomma 
\end{equation}
where we introduced $\hat{\vect{e}}$ as the unit vector along the direction of the mean-field order parameter. For CDW~order, $\hat{\vect{e}}=1$, while for SDW~order, we assume w.l.o.g.\ $\hat{\vect{e}}=\hat{\vect{e}}_3$ in the remainder. 

The usual procedure of integrating out the fermions after the decoupling~\eqref{eq:decoupling} then leads to the effective action in terms of the fermionic Green's function
 \begin{equation}
 \mathcal{S}_\mathrm{eff}(\Phi_0)=\int_q\frac{2|\vect{\Phi}_q|^2}{V}-\int_{k,k^\prime}\tr\ln(\mathcal{G}_{kk^\prime}^{-1}) \mathcomma 
\end{equation}
which can be split into a mean-field part and fluctuations as well using 
\begin{align}
&\tr\ln\big(-\mathcal{G}^{-1}_{kk^\prime}\big)=\tr\ln\big[-\big((\mathcal{G}^\mathrm{MF}_k)^{-1}\delta_{kk^\prime}+\eta_{kk^\prime}\big)\big]\\
&\quad =\tr\ln\big[-\big(\mathcal{G}^{\mathrm{MF}}_k)^{-1}\delta_{kk^\prime}\big)\big]-\frac{1}{2}\tr\big(\mathcal{G}^\mathrm{MF}_k\eta_{kk^\prime}\mathcal{G}^\mathrm{MF}_{k^\prime}\eta_{k^\prime k}\big) \nonumber \\ 
&\quad \qquad +\mathcal{O}\big[(\delta\vect{\Phi})^3\big] 
\end{align}
by expanding the fluctuations up to Gaussian order. 
Here, the mean-field part of the inverse matrix Green's function in Matsubara representation is given by 
\begin{align}
 &\big(\mathcal{G}_k^{\mathrm{MF}})^{-1}=\tfrac{1}{2}\left(\mathrm{i}\nu_n\tau_0\sigma_0-\varepsilon_{\vect{k}}\tau_3\sigma_0\right) \nonumber \\ 
  &\qquad +\tfrac{1}{2}\tau_1\left\{\begin{matrix}\Phi_0 \sigma_0 & \text{for CDW order} \mathcomma\\ \vect{\Phi}_0\cdot \vect{\sigma}& \text{for SDW order}\mathcomma\end{matrix}\right.
\end{align}
while the fluctuation matrix is given by 
\begin{equation}
 \eta_{kk^\prime}=\tfrac{1}{2}\tau_1\left\{\begin{matrix} \delta\Phi_{k-k^\prime}\sigma_0 & \text{for CDW order} \mathcomma \\ \delta\vect{\Phi}_{k-k^\prime}\cdot\vect{\sigma} & \text{for SDW order}\mathperiod \end{matrix} \right. 
\end{equation}
Note that when considering density-wave order with an imaginary order parameter, the order parameter and its fluctuations are associated with $\tau_2$ rather than $\tau_1$. One easily finds that this change will not affect the conclusions of our analysis. 
Then the partition function can be expressed as 
\begin{equation}
 \mathcal{Z}=\int\mathcal{D}[\bar{\Psi},\Psi]\, \mathrm{e}^{-(\mathcal{S}_0+\mathcal{S}_\mathrm{int})}\approx\mathrm{e}^{-(\mathcal{S}_\mathrm{MF}+\mathcal{S}_\mathrm{fluct})} \mathcomma 
\end{equation}
where the mean-field action takes the form 
\begin{equation}
 \mathcal{S}_\mathrm{MF}(\Phi_0)=\frac{2\Phi_0^2}{V}-\int_k\tr\ln\Big[-\big(\mathcal{G}_k^\mathrm{MF}\big)^{-1}\Big]\mathcomma 
\end{equation}
resulting in the famous form of the gap equation 
\begin{align}
 0&=\frac{\dd\mathcal{S}_\mathrm{MF}(\Phi_0)}{\dd\Phi_0^2}=\frac{2}{V}-\int\frac{\dd^d\vect{k}}{(2\pi)^d}\int_{-\infty}^\infty\frac{\dd\nu}{2\pi}\,\frac{2}{\varepsilon_{\vect{k}}^2+\nu^2+\Phi_0^2} \nonumber \\ 
 &\approx\frac{2}{V}-2\dosf\ln\Big(\frac{2 E_\mathrm{F}}{\Phi_0}\Big) 
\end{align}
at zero temperature. 
The Gaussian fluctuation part can be integrated exactly, 
\begin{align}\begin{split}
 \mathrm{e}^{-\mathcal{S}_\mathrm{fluct}}&=\int\mathcal{D}[\delta\vect{\Phi}]\,\mathrm{e}^{-\frac{1}{2}\int_q\delta\vect{\Phi}_q\cdot\vect{\mathcal{D}}_q^{-1}\cdot\delta\vect{\Phi}_{-q}} \\  
 &= \mathrm{e}^{-\frac{1}{2}\int_q\ln\det[\vect{\mathcal{D}}^{-1}_q]}\mathperiod
\end{split}\end{align}
Alternatively, the inverse fluctuation propagator matrix~$\vect{\mathcal{D}}^{-1}_q$, and consequently the Gaussian fluctuation corrections to the action, can be expressed in terms of the polarization matrix~$\vect{\Pi}_q$ as 
\begin{equation}
\vect{\mathcal{D}}_q^{-1}= \frac{4}{V}\vect{\mathds{1}}-\vect{\Pi}_q
\end{equation}
where $\vect{\Pi}_q$ is either given by 
\begin{equation}
 \Pi^\mathrm{CDW}_q=-\frac{1}{4}\int_k\tr\big[\mathcal{G}^\mathrm{CDW}_{k+\frac{q}{2}}(\tau_1\sigma_0)\mathcal{G}^\mathrm{CDW}_{k-\frac{q}{2}}(\tau_1\sigma_0)\big] \mathcomma 
\end{equation}
or by 
\begin{align}
(\vect{\Pi}^\mathrm{SDW}_q)_{ij}&=-\frac{1}{4}\int_k\tr\big[\mathcal{G}^\mathrm{SDW}_{k+\frac{q}{2}}(\tau_1\sigma_i)\mathcal{G}_{k-\frac{q}{2}}^\mathrm{SDW}(\tau_1\sigma_j)\big] \\
 &=\begin{pmatrix} 4\Pi^\perp_q & 0 &0 \\ 0 &  4\Pi^\perp_q & 0\\ 0 & 0 & 4 \Pi^\parallel_q\end{pmatrix}\label{eq:Pi-SDW-matrix}\mathperiod
\end{align}
In the last line, we introduced longitudinal and transverse contributions 
\begin{align}
 \Pi^\perp_q&=\Pi^\mathrm{n}_q-\Pi^\mathrm{a}_q  \\
 \Pi^\parallel_q&=\Pi^\mathrm{n}_q+\Pi^\mathrm{a}_q 
\end{align}
in terms of the normal~($\Pi_q^\mathrm{n}$) and anomalous~($\Pi_q^\mathrm{a}$) part of the polarization function, which are discussed in more detail in Appendix~\ref{app:polarization-function}. 
The generalization to spin dimensionality~$N$ is straightforward, and the resulting polarization matrix differs from Eq.~\eqref{eq:Pi-SDW-matrix} only in the number of transverse modes. The fluctuation corrections to the action then take the form 
\begin{equation}
 \mathcal{S}_\mathrm{fluct}(\Phi_0)=\frac{1}{2}\int_q\ln\Big[4^N\Big(\frac{1}{V}-\Pi_q^\perp\Big)^{N-1}\Big(\frac{1}{V}-\Pi_q^\parallel\Big)\Big]
\end{equation}
for both SDW~order and CDW~order, where the latter corresponds to~$N=1$. 

Owing to the diagonal structure of~$\vect{\mathcal{D}}^{-1}_q$, the Gaussian fluctuation corrections to the gap equation, 
\begin{align} 
 \frac{\dd \mathcal{S}_\mathrm{fluct}(\Phi_0)}{\dd\Phi_0^2}&=\frac{1}{2}\int_q\frac{\frac{\dd}{\dd\Phi_0^2}\det(\vect{\mathcal{D}}_q^{-1})}{\det(\vect{\mathcal{D}}_q^{-1})} \\ 
 &= -\frac{1}{2}\int_q\Bigg[\frac{(N-1)\frac{\dd\Pi^\perp_q}{\dd\Phi_0^2}}{\frac{1}{V}-\Pi^\perp_q}+\frac{\frac{\dd\Pi^\parallel_q}{\dd\Phi_0^2}}{\frac{1}{V}-\Pi^\parallel_q}\Bigg]\mathcomma \nonumber
\end{align}
can alternatively be written as 
\begin{equation}
 \frac{\dd \mathcal{S}_\mathrm{fluct}(\Phi_0)}{\dd\Phi_0^2}=-\frac{1}{2}\int_q 
 \Big(\vect{\mathcal{D}}_q\cdot\frac{\dd\vect{\Pi}_q}{\dd\Phi_0^2}\Big)\mathcomma \label{eq:fluctuation-integral}
\end{equation}
which is represented by the diagram in Fig.~\ref{fig:gap-equation}(b), where the fermionic triangle (also see Fig.~\ref{fig:triangles}) represents the derivative of the polarization function, which is also evaluated in Appendix~\ref{app:polarization-function}. 
As shown in Refs.~\onlinecite{KosMillisLarkin-PRB2004} and~\onlinecite{FischerEtAl2017}, the integral~\eqref{eq:fluctuation-integral} is dominated by contributions from the regime where $r=\sqrt{\omega^2+(v_\mathrm{F}|\vect{q}|\cos\theta)^2}/(2\Phi_0)\gg1$ as well as $v_\mathrm{F}|\vect{q}|>\omega$, whereas corrections stemming from long-wavelength fluctuations are negligible. 
In that regime, the fluctuation propagator is dominated by the normal part of the polarization function, $|1/V-\Pi^\mathrm{n}_q|\gg|\Pi^\mathrm{a}_q|$, while the leading contribution to the derivative of the polarization function stems from the anomalous part, $|\dd\Pi^\mathrm{a}_q/\dd\Phi_0^2|\gg|\dd\Pi^\mathrm{n}_q/\dd\Phi_0^2|$. Consequently, the crucial correction terms to the gap equation due to fluctuations are given by 
\begin{align}
 \left.\frac{\dd\mathcal{S}_\mathrm{fluct}(\Phi_0)}{\dd\Phi_0^2}\right|_{r\gg1}\approx \frac{1}{2}\int_q\frac{[(N-1)-1]\frac{\dd\Pi_q^\mathrm{a}}{\dd\Phi_0^2}}{\frac{1}{V}-\Pi_q^\mathrm{n}}\mathperiod 
\end{align}
Then, depending on the number of transverse modes, i.\,e., the dimensionality of the order parameter, the fluctuation corrections are either positive ($N<2$), negative ($N>2$), or negligible ($N=2$). 
\section{The polarization function and its derivatives}\label{app:polarization-function}
For the sake of completeness, we summarize the results for the polarization function and its derivatives obtained by previous studies~\cite{KosMillisLarkin-PRB2004,FischerEtAl2017}. 
To pinpoint the physical meaning of the individual terms eventually contributing to the gap equation, we introduce normal and anomalous fermionic Green's functions 
as 
\begin{align}
 G_k&\equiv G(\vect{k},\nu_n)= -\frac{\mathrm{i}\nu_n+\varepsilon_{\vect{k}}}{\varepsilon_{\vect{k}}^2+\nu_n^2+\Phi_0^2} \\ 
 \text{and} \quad F_k&\equiv F(\vect{k},\nu_n)= \frac{\Phi_0}{\varepsilon_{\vect{k}}^2+\nu_n^2+\Phi_0^2}\mathcomma 
\end{align}
respectively. It holds that $F_k=F_{-k}$, and furthermore, nesting implies that $G(\vect{k}+\vect{Q},\nu_n)=-G(-\vect{k},-\nu_n)\equiv -G_{-k}$ as well as $F(\vect{k}+\vect{Q},\nu_n)=F(\vect{k},\nu_n)\equiv F_k$.  Using these definitions, the matrix Green's functions can be rewritten as 
\begin{align}
 \mathcal{G}_k^\mathrm{CDW}&= 2\begin{pmatrix} G_k\sigma_0 & F_k\sigma_0 \\ F_k\sigma_0 & -G_{-k}\sigma_0 \end{pmatrix} \\ 
 \text{and} \quad \mathcal{G}_k^\mathrm{SDW} &= 2\begin{pmatrix} G_k \sigma_0 & F_k \sigma_3 \\ F_k\sigma_3 & -G_{-k}\sigma_0\end{pmatrix} \mathcomma
\end{align}
which makes the structural congruence with the BCS~theory of superconductivity obvious. 
The normal part of the polarization function is then solely determined by normal Green's functions. It corresponds to $\Pi^\mathrm{n}_q=\frac{1}{2}(\Pi_{11,q}+\Pi_{22,q})$ in the notation of Ref.~\onlinecite{FischerEtAl2017} and reads 
\begin{subequations}
\begin{align}
 \Pi^\mathrm{n}_q&= \frac{1}{2}\int_k\big[G_{k+\frac{q}{2}}G_{-(k-\frac{q}{2})}+G_{-(k+\frac{q}{2})}G_{k-\frac{q}{2}}\big] \\
 &=\int_k\frac{\nu_+\nu_-+\varepsilon_+\varepsilon_-}{(\varepsilon_+^2+\nu_+^2+\Phi_0^2)(\varepsilon_-^2+\nu_-^2+\Phi_0^2)} \\
 &=\dosf\ln\bigg(\frac{2 E_\mathrm{F}}{\Phi_0}\bigg)-\dosf\int_\Omega\frac{(2r^2+1)\operatorname{arsinh}(r)}{2r\sqrt{r^2+1}}\label{eq:normal-Pi}  \mathcomma 
 \end{align}\end{subequations}
where we abbreviated~$\int_k\ldots=\dosf\int\dd\varepsilon\int\frac{\dd\nu}{2\pi}\int_\Omega\ldots$ and the integration $\int_\Omega\dots =\frac{1}{\Omega_d}\int\dd\Omega\dots$ refers to the integration over the direction of the fermionic momentum with $\Omega_d$ being the volume of a $d$-dimensional sphere. Since we consider the spectrum to be isotropic, the integrand only depends on the momentum direction via $\theta=\sphericalangle(\vect{k},\vect{q})$. Furthermore, we introduced~$\nu_\pm=\nu\pm\frac{\omega}{2}$ as well as~$\vect{k}_\pm=\vect{k}\pm\frac{1}{2}\vect{q}$, and linearized the dispersion~$\varepsilon_\pm=\varepsilon\pm \frac{1}{2}v_\mathrm{F}|\vect{q}|\cos\theta$. 
For the evaluation of the polarization function, it is useful that the integrals discussed here depend on external momenta and frequency only via the combination 
\begin{equation}
 r=\frac{\sqrt{\omega^2+(v_\mathrm{F} |\vect{q}|\cos\theta)^2}}{2\Phi_0}\mathcomma 
\end{equation}
see Refs.~\onlinecite{KosMillisLarkin-PRB2004} and~\onlinecite{VaksGalitskiiLarkin-JETP1962} for details. Evaluating the above expression at $|\vect{q}|=0$ and $\omega=2\Phi_0r$ then results in the last line, where the angular integration still remains to be done. Unfortunately, this cannot be performed for arbitrary values of~$r$, and we resort to approximations in the regimes~$r\ll1$ and $r\gg1$, cf.\ Refs.~\onlinecite{KosMillisLarkin-PRB2004} and~\onlinecite{FischerEtAl2017} for details. 

Analogously, the anomalous part of the polarization function can be expressed as 
 \begin{subequations}
 \begin{align}
 \Pi^\mathrm{a}_q&= -\int_k F_{k+\frac{q}{2}}F_{k-\frac{q}{2}}  \\ 
 &= -\int_k\frac{\Phi_0^2}{(\varepsilon_+^2+\nu_+^2+\Phi_0^2)(\varepsilon_-^2+\nu_-^2+\Phi_0^2)} \\ 
 &= -\dosf\int_\Omega\frac{\operatorname{arsinh}(r)}{2r\sqrt{r^2+1}} \label{eq:anomalous-Pi} \mathperiod
\end{align}\end{subequations}
In the notation of Ref.~\onlinecite{FischerEtAl2017}, this corresponds to $\Pi^\mathrm{a}_q=\frac{1}{2}(\Pi_{11,q}-\Pi_{22,q})$. 
This part vanishes in the limit~$\Phi_0\rightarrow0$, i.\,e., in the disordered high-temperature phase. 
Furthermore, it is obvious from the expressions~\eqref{eq:normal-Pi} and~\eqref{eq:anomalous-Pi} that
\begin{equation}
 \left|\tfrac{1}{V}-\Pi_q^\mathrm{n}\right|>\left|\Pi_q^\mathrm{a}\right|
\end{equation}
holds for arbitrary~$r$. 

\begin{figure}[h]
 \includegraphics[width=\columnwidth]{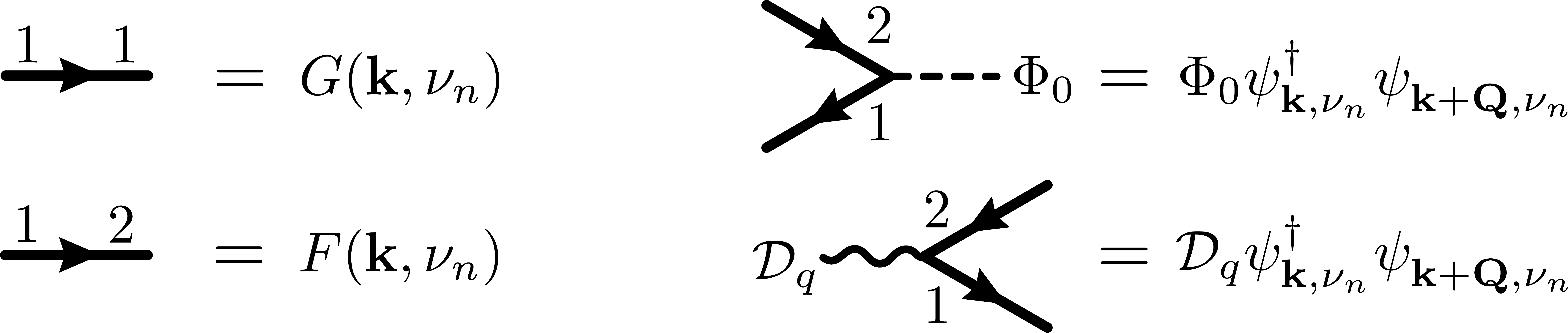}
 \caption{Diagrammatic elements. The normal and anomalous propagators are represented by straight lines, and the numbers refer to the matrix structure in the band space emerging as a consequence of the doubling of the unit cell, as introduced in Eq.~\eqref{eq:spinor}. Furthermore, two types of vertices appear in the diagrammatic representation of the gap equation: the coupling to the order parameter as well as to the respective fluctuations.}
 \label{fig:diagrammatic-elements}
\end{figure}

The corresponding derivatives can also be expressed in terms of normal and anomalous Green's function, and the individual terms are represented by the fermionic triangle diagrams shown in Fig.~\ref{fig:triangles}. All contributions to the derivative of the normal part, $\frac{\dd\Pi_q^\mathrm{n}}{\dd\Phi_0^2}=-\frac{1}{2}\Big(\frac{\dd\mathcal{D}^{-1}_{11,q}}{\dd\Phi_0^2}+\frac{\dd\mathcal{D}^{-1}_{22,q}}{\dd\Phi_0^2}\Big)$,  have the structure presented in Fig.~\ref{fig:triangles}(c), and the analytic expression is given by 
\begin{widetext}
\begin{subequations}
\begin{align}
 \frac{\dd\Pi_q^\mathrm{n}}{\dd\Phi_0^2}&=-\frac{1}{2\Phi_0}\int_k\big[G_{k+\frac{q}{2}}F_{k+\frac{q}{2}}G_{-(k-\frac{q}{2})}+G_{k+\frac{q}{2}}F_{-(k-\frac{q}{2})}G_{-(k-\frac{q}{2})}+G_{k-\frac{q}{2}}F_{k-\frac{q}{2}}G_{-(k+\frac{q}{2})}+G_{k-\frac{q}{2}}F_{-(k+\frac{q}{2})}G_{-(k+\frac{q}{2})}\big] \\
 &=-\frac{\dosf}{4\Phi_0^2}\int_\Omega\bigg[\frac{\operatorname{arsinh}(r)}{r(r^2+1)^{3/2}}+\frac{1}{r^2+1}\bigg]\mathperiod
\end{align}
\end{subequations}
The derivation of~$\Pi_q^\mathrm{a}$ w.r.t.~$\Phi_0^2$ generates two different types of contributions -- the first [cf.\ Fig.~\ref{fig:triangles}(a)] comes from the exchange of fluctuations between quasiparticles  while the second [cf.\ Fig.~\ref{fig:triangles}(b)] is solely determined by anomalous propagators, 
\begin{subequations}
\begin{align}
 \frac{\dd\Pi_q^\mathrm{a}}{\dd\Phi_0^2}&=-\frac{1}{2\Phi_0}\int_k\big[G_{k+\frac{q}{2}}G_{-(k+\frac{q}{2})}F_{k-\frac{q}{2}}+G_{k-\frac{q}{2}}G_{-(k-\frac{q}{2})}F_{k+\frac{q}{2}}-F_{k+\frac{q}{2}}F_{k+\frac{q}{2}}F_{k-\frac{q}{2}}-F_{k+\frac{q}{2}}F_{k-\frac{q}{2}}F_{k-\frac{q}{2}}\big]\\
 &=-\frac{\dosf}{4\Phi_0^2}\int_\Omega\bigg[\frac{(2r^2+1)\operatorname{arsinh}(r)}{r(r^2+1)^{3/2}}-\frac{1}{r^2+1}\bigg]\mathperiod
\end{align}
\end{subequations}
\end{widetext}
The crucial contribution to $\frac{\dd\Pi_q^\mathrm{a}}{\dd\Phi_0^2}=-\frac{1}{2}\Big(\frac{\dd\mathcal{D}^{-1}_{11,q}}{\dd\Phi_0^2}-\frac{\dd\mathcal{D}^{-1}_{22,q}}{\dd\Phi_0^2}\Big)$ that ultimately generates the additional logarithm contributing to the gap equation stems from the term~$\propto\int_\Omega \ln(2r)/r^2$ in $\dd\Pi_q^\mathrm{a}/\dd\Phi_0^2$ in the regime $r\gg1$. It can be related to the first term, i.\,e., the contribution depicted in Fig.~\ref{fig:triangles}(a). The diagrammatic key elements used throughout the paper are introduced in Fig.~\ref{fig:diagrammatic-elements}.

\section{Effect of channel interference on density-wave instabilities} \label{app:two-band}
In this appendix, we demonstrate how the presence of competing instabilities can affect the transition towards a new low-temperature ordered phase. 
We use the two-band model of iron-based superconductors as an example, which has been studied in great detail in the recent past~\cite{FernandesChubukov-RepProgPhys2017}. 
In our brief analysis, we greatly build on the RG~analysis of this model as presented in Ref.~\onlinecite{ChubukovEfremovEremin-PRB2008}. We start with a brief summary of their results before using them to analyze the effect of channel interference on density-wave instabilities. 

The model analyzed in this appendix consists of two nested Fermi pockets: one with a hole-like dispersion centered around~$\vect{0}$, and another one with an electron-like dispersion centered around~$\vect{Q}$, which we assume to be perfectly nested. 
The notation used in this appendix then merely differs from the notation used in the main text in that we introduce the band index~$j\in\{1,2\}$ and measure momenta~$\vect{k}$ as deviations from~$\vect{0}$ and~$\vect{Q}$, respectively. 
The noninteracting part~[cf.\ Eq.~\eqref{eq:H0}] of the Hamiltonian~$\mathcal{H}=\mathcal{H}_0+\mathcal{H}_\mathrm{int}$ then reads 
\begin{equation} 
 \mathcal{H}_0= \sum_{j,\vect{k},\sigma}\varepsilon_{j,\vect{k}}\psi^\dagger_{j,\vect{k},\sigma}\psi_{j,\vect{k},\sigma}\mathcomma
\end{equation}
and the nesting condition takes the form $\varepsilon_{1,\vect{k}}=-\varepsilon_{2,\vect{k}}$. 
Furthermore, the interaction~\eqref{eq:H_interaction} translated to band notation is given by 
\begin{align}
 \mathcal{H}_\mathrm{int}&= \frac{1}{2}\sum\psi^\dagger_{j_1,\vect{k}_1,\sigma_1}\psi^\dagger_{j_2,\vect{k}_2,\sigma_2} U^{\sigma_1\sigma_2,\sigma_3\sigma_4}_{j_1j_2,j_3j_4}(\vect{k}_1,\vect{k}_2;\vect{k}_3,\vect{k}_4) \nonumber \\ &\qquad \times  \psi^{}_{j_3,\vect{k}_3,\sigma_3}\psi^{}_{j_4,\vect{k}_4,\sigma_4}\delta(\vect{k}_1+\vect{k}_2-\vect{k}_3-\vect{k}_4)\mathcomma \label{eq:H_int}
\end{align}
where the summation is over band indices~$j_i$, momenta~$\vect{k}_i$, and spins~$\sigma_i$. 
In contrast to the discussion in the main text, we allow for a weak momentum dependence of the couplings in the sense that they still depend on band indices, i.\,e., on whether the momenta are close to~$\vect{0}$ or close to~$\vect{Q}$. This results in several coupling constants associated with the different intraband and interband processes. 
After decomposing the interaction into charge~(ch) and spin~(sp) channel according to
\begin{align}
 U^{\sigma_1\sigma_2,\sigma_3\sigma_4}_{j_1j_2,j_3j_4}(\vect{k}_1\vect{k}_2;\vect{k}_3,\vect{k}_4)= U_{j_1j_2;j_3j_4}^\mathrm{ch}\delta_{\sigma_1\sigma_4}\delta_{\sigma_2\sigma_3} \nonumber \\  + U_{j_1j_2;j_3j_4}^\mathrm{sp}\vect{\sigma}_{\sigma_1\sigma_4}\cdot\vect{\sigma}_{\sigma_2\sigma_3}\mathcomma
\end{align}
the spin sums can be partially evaluated and the resulting expression for the interaction term contains five independent interaction constants $U_i^{(0)}$, 
\begin{align}
 \mathcal{H}^\prime_\mathrm{int}&= U_1^{(0)}\sum_{\vect{k}\vect{k}^\prime\vect{q}}\psi^\dagger_{1,\vect{k},\sigma}\psi^\dagger_{2,\vect{k}^\prime,\sigma^\prime}\psi^{}_{2,\vect{k}^\prime-\vect{q},\sigma^\prime}\psi^{}_{1,\vect{k}+\vect{q},\sigma} \nonumber  \\ 
 &\quad + U_2^{(0)}\sum_{\vect{k}\vect{k}^\prime\vect{q}}\psi^\dagger_{2,\vect{k},\sigma}\psi^\dagger_{1,\vect{k}^\prime,\sigma^\prime}\psi^{}_{2,\vect{k}^\prime-\vect{q},\sigma^\prime}\psi^{}_{1,\vect{k}+\vect{q},\sigma} \nonumber \\
 &\quad + \frac{U_3^{(0)}}{2}\sum_{\vect{k}\vect{k}^\prime\vect{q}}\big[\psi^\dagger_{2,\vect{k},\sigma}\psi^\dagger_{2,\vect{k}^\prime,\sigma^\prime}\psi_{1,\vect{k}^\prime-\vect{q},\sigma^\prime}\psi_{1,\vect{k}+\vect{q},\sigma}+\mathrm{H.\,c.}\big] \nonumber  \\ 
 &\quad + \frac{U_4^{(0)}}{2}\sum_{\vect{k}\vect{k}^\prime\vect{q}}\psi^\dagger_{2,\vect{k},\sigma}\psi^\dagger_{2,\vect{k}^\prime,\sigma^\prime}\psi^{}_{2,\vect{k}^\prime-\vect{q},\sigma^\prime}\psi^{}_{2,\vect{k}+\vect{q},\sigma} \nonumber \\
 &\quad + \frac{U_5^{(0)}}{2}\sum_{\vect{k}\vect{k}^\prime\vect{q}}\psi^\dagger_{1,\vect{k},\sigma}\psi^\dagger_{1,\vect{k}^\prime,\sigma^\prime}\psi^{}_{1,\vect{k}^\prime-\vect{q},\sigma^\prime}\psi^{}_{1,\vect{k}+\vect{q},\sigma}\mathcomma 
 \label{eq:H_int2}
\end{align}
which we labeled in accordance with Ref.~\onlinecite{ChubukovEfremovEremin-PRB2008}.  
Here the index~$0$ indicates that these are the bare couplings. 
Note that the interband pair-hopping process associated with $U_3^{(0)}$ is only allowed if~$2\vect{Q}$ is from the reciprocal lattice, as it is the case for iron-based superconductors. 
$U_4^{(0)}$ and $U_5^{(0)}$ are the intraband couplings in the two bands, and as we consider the system at perfect nesting, i.\,e., particle-hole symmetry, it holds that~$U_4^{(0)}=U_5^{(0)}$. $U_1^{(0)}$ and $U_2^{(0)}$ refer to interband processes with a momentum transfer of~$\vect{0}$ and~$\vect{Q}$, respectively.
The couplings~$U_i^{(0)}$ are connected to the couplings in spin and charge channel by 
\begin{subequations}\label{eq:couplings-connection}
\begin{align}
 U_{11,11}^\mathrm{ch}&= \frac{U_5^{(0)}}{4} \mathcomma & U_{11,11}^\mathrm{sp}&=-\frac{U_5^{(0)}}{4} \mathcomma \\ 
 U_{22,22}^\mathrm{ch}&= \frac{U_4^{(0)}}{4} \mathcomma & U_{22,22}^\mathrm{sp}&=-\frac{U_4^{(0)}}{4} \mathcomma \\
 U_{11,22}^\mathrm{ch}&= \frac{U_3^{(0)}}{4} \mathcomma  & U_{11,22}^\mathrm{sp}&=-\frac{U_3^{(0)}}{4} \mathcomma \\
 U_{12,12}^\mathrm{ch}&= -\frac{U_1^{(0)}}{4}+\frac{U_2^{(0)}}{2} \mathcomma  & U_{12,12}^\mathrm{sp}&=-\frac{U_1^{(0)}}{4} \mathcomma  \\ 
 U_{12,21}^\mathrm{ch}&= \frac{U_1^{(0)}}{2}-\frac{U_2^{(0)}}{4} \mathcomma & U_{12,21}^\mathrm{sp}&=-\frac{U_2^{(0)}}{4} \mathperiod
\end{align}
\end{subequations}
In the remainder, we will work with dimensionless quantities, and to this end, we introduce dimensionless couplings via $u_i=\dosf U_i$, where $\dosf$ is the density of states at the Fermi level. 

The presence of the interaction~\eqref{eq:H_int} implies three different types of instabilities: towards the formation of charge-density wave~(CDW) order and spin-density wave~(SDW) order, both with momentum~$\vect{Q}$, as well as a superconducting~(SC) instability resulting from Cooper pairing either in the conventional~$s^{++}$-wave~channel or in the sign-changing~$s^{+-}$~channel. 
The couplings~$\Gamma$ in the respective channels [cf.\ the couplings $V_\mathrm{ch}$ and $V_\mathrm{sp}$ as introduced in Eq.~\eqref{eq:interaction} in the main text] are given by the combinations 
\begin{subequations}\label{eq:couplings}
\begin{align}
 \Gamma_\mathrm{rSDW}&= u_1+ u_3 \mathcomma & \Gamma_\mathrm{iSDW}&= u_1-u_3 \mathcomma\\ 
 \Gamma_\mathrm{iCDW}&= u_1+u_3-2u_2 \mathcomma & \Gamma_\mathrm{rCDW}&= u_1-u_3-2u_2 \mathcomma\\ 
 \Gamma_{s^{+-}}&=u_4-u_3 \mathcomma & \Gamma_{s^{++}}&= u_4+u_3 \mathperiod
 \end{align}
 \end{subequations}
Here, the labels~r and~i for density-wave instabilities refer to density-wave order characterized by a purely real and a purely imaginary order parameter, respectively. 
What is more, the interaction~\eqref{eq:H_int} possesses an $\mathrm{SO}(6)$~symmetry provided that $u_2^{(0)}=0$ and $u_4^{(0)}=-u_1^{(0)}$, as discussed by Ref.~\onlinecite{PodolskyKeeKim-EPL2009}. 
Then, three out of the six states of emerging order are degenerate in energy -- namely rSDW, iCDW, and $s^{+-}$~SC for repulsive interband pair hopping~($u_3^{(0)}>0$), whereas attractive interband pair hopping~($u_3^{(0)}<0$) may result in iSDW, rCDW, and $s^{++}$~SC. 
Real materials only approximately exhibit this enhanced symmetry, meaning that one of the instabilities wins. In the parent compounds of iron-based superconductors, for instance, spin-density wave order is realized and hence the coupling in the rSDW~channel is the one which diverges first upon successively integrating out high-energy modes in a renormalization group~(RG) analysis of the model. However, since other candidates for low-energy ordered phases are close in energy, they are competing for the same electrons, implying that phase competition is important in such systems. 
In the remainder, we analyze the effect of channel interference on the energy scale at which the instability towards density-wave order occurs. 

The RG~flow of the couplings $u_i$ as functions of~$t=\log\tfrac{W}{E}$, where~$W$ is the bandwidth and $E$ the running energy scale, is governed by the coupled differential equations 
\begin{subequations}\label{eq:RG-equations}
\begin{align}
 \dot{u}_1&= u_1^2+u_3^2 \mathcomma \\ 
 \dot{u}_2&= 2u_2(u_1-u_2) \mathcomma \\
 \dot{u}_3&= 2u_3(2u_1-u_2-u_4) \mathcomma \\ 
 \dot{u}_4&= -u_3^2-u_4^2 \mathperiod
\end{align}
\end{subequations}
For the derivation and a detailed discussion, we refer to Ref.~\onlinecite{ChubukovEfremovEremin-PRB2008}. Analogous results have been obtained by Refs.~\onlinecite{Schlottmann-PRB1999,Schlottmann-PRB2003} for a related model without the pair-hopping process. 
Consequently, the flow of the couplings in the density-wave channels takes the form 
\begin{subequations}\label{eq:channel-interference}
 \begin{align}
  \dot{\Gamma}_\mathrm{SDW}&= (\Gamma_\mathrm{SDW})^2\pm 2u_3(u_1-u_2-u_4)\mathcomma \\ 
  \dot{\Gamma}_\mathrm{CDW}&= (\Gamma_\mathrm{CDW})^2\mp 2u_3(u_1+u_2-u_4) \mathcomma 
 \end{align}
 \end{subequations}
where the upper sign refers to rSDW and rCDW, while the lower sign refers to iSDW and iCDW. If the second term were zero, this would result in the usual logarithmic divergence~$\Gamma=\Gamma_0/(1-\Gamma_0\log\frac{W}{E})$ of the coupling in the respective channel. Hence the presence of the second term implies corrections due to channel interference, meaning that these effects are intrinsic to the model. They vanish only for $u_3^{(0)}=0$, which we cannot generically assume for the iron-based superconductors as $2\vect{Q}$ is a reciprocal lattice vector here. Let us further note here that the second term combines the effect of competing density-wave and superconducting instabilities, even though we cannot differentiate between the effect of different channels on a given instability within this approach. 

Motivated by Eq.~\eqref{eq:channel-interference}, we use $u_3^{(0)}$ as a measure of channel-interference strength. We may then analyze whether channel interference is beneficial or detrimental to the formation of a certain type of order by numerically solving the RG~equations~\eqref{eq:RG-equations} for different values of~$u_3^{(0)}$. 
For notational convenience, let us concentrate on the parameter range appropriate to describe the physics of iron-based superconductors here, i.\,e., repulsive interband pair hopping $u_3^{(0)}>0$ leading to the competition of~rSDW~order with iCDW~order and $s^{+-}$~SC. We note here that the same trends are found mutatis mutandis for attractive $u_3^{(0)}<0$, i.\,e., for competing iSDW, rCDW, and $s^{++}$~SC~instabilities.  
In order to analyze the effect of channel interference, we compare the flow of the coupling in a given channel for fixed $\Gamma^{(0)}$ upon varying $u_3^{(0)}$, which constitutes a measure of channel interference strength. 
Here, the bare parameters~$u_i^{(0)}$ are chosen such that the ordering in the channel under consideration is favorable within the mean-field description, that is, $\Gamma^{(0)}\neq0$, while the bare couplings in the competing channels are tuned to zero. 
Although the couplings in the competing channels grow with the flow and are relevant as well, the leading instability remains the same as long as the channel interference does not change which of the couplings diverges first. However, the energy scale at which the couplings diverge turns out to be affected by the bare value of the interband pair hopping~$u_3^{(0)}$, which can be used as a sign indicating whether channel interference promotes or hinders the formation of order in a given channel. 
In Fig.~\ref{fig:channel-interference}(a), we exemplarily show the flow of $\Gamma_\mathrm{rSDW}$ obtained from solving Eqs.~\eqref{eq:RG-equations} for fixed bare interaction~$\Gamma_\mathrm{rSDW}^{(0)}=0.5$. 
Upon increasing the channel interference strength via increasing~$u_3^{(0)}$, the energy scale at which the interaction in the rSDW channel diverges is pushed to lower energies, that is, happens at higher transition temperatures. 
As a result, we find that channel interference is beneficial for the formation of SDW~order. 
On the other hand, the energy scale at which the interaction~$\Gamma_\mathrm{iCDW}$ in the iCDW~channel diverges is pushed to higher energies upon increasing the channel interference strength~$u_3^{(0)}$ while keeping $\Gamma_\mathrm{iCDW}^{(0)}=0.5$ fixed. Therefore, channel interference is detrimental to the formation of CDW~order as the phase transition now happens at lower temperatures, as illustrated in Fig.~\ref{fig:channel-interference}(b). 

%

\end{document}